# Topological phase dynamics described by overtone-synthesized classical and quantum Adler equations

Hiroshi Yamaguchi[*] and Motoki Asano

Basic Research Laboratories, NTT, Inc., Atsugi, Kanagawa, 243-0198 Japan

[*]*hrsh.yamaguchi@ntt.com*

The Adler equation is a well-known one-dimensional model describing phase locking and synchronization. Motivated by recent experiments using optomechanical oscillators, we extend the model to include overtone-synthesized sinusoidal coupling with adiabatic temporal modulation. This extension gives rise to unique topological features such as winding-number quantization, discontinuous phase-slip transitions, and hysteretic and non-reciprocal phase dynamics. We further extend the analysis to the quantum regime, where we find a counterintuitive result: the breakdown of winding-number quantization. This arises from the superposition of different winding-number states in a closed-space Thouless pump. Moreover, hysteretic dynamics, once eliminated in quantum adiabatic approximation, is recovered in non-adiabatic calculations, as the superposition of two Floquet states with different PT eigenvalues becomes the quantum counterpart of phase trajectory.

## I. INTRODUCTION

Phase locking and synchronization are commonly observed phenomena in various oscillator systems [1-11]. The dynamics are described by an Adler equation [2], which models the oscillator coupling as a sinusoidal function of the phase difference. This behavior extends beyond ensembles of oscillators, described by the Kuramoto model, a multimode extension of the Adler equation [12-14]. While these models assume a single-tone sinusoidal and time-independent coupling, generalizations including overtone-synthesized couplings and temporal modulation have been proposed [15-17] and experimentally implemented [18-20]. In particular, micro-bottle optomechanical systems [20] enable the overtone-synthesized coupling, as well as the temporal control of the coupling parameters, via optical intensity modulation. They exhibit distinctive phenomena such as hysteretic phase slips, discontinuous winding-number transitions, and non-reciprocal phase-slip dynamics, highlighting topological features arising from time-dependent generalized couplings.

The system with a periodic and adiabatic coupling-potential modulation is analogous to the Thouless pump, introduced in 1983 to incorporate the concept of Berry curvature into a one-dimensional quantum system by treating time as an additional adiabatic parameter [21-22]. From the requirement on the wave function to be single-valued, integrating the Berry curvature over the 1+1-dimensional parameter space yields a quantized current flow per modulation period. This concept, once purely theoretical, has recently been demonstrated in real systems such as ultracold atoms [23-24], optical waveguides [25-27], and acoustic waveguides [28].

In this study, we connect these two independent concepts—classical synchronization and the quantum Thouless pump—by investigating phase dynamics in temporally modulated classical and quantum overtone-synthesized Adler equations. The two-tone coupling in this model is periodically modulated, leading to quantized winding numbers, reminiscent of the Thouless pump. A key distinction, however, lies in their variable spaces; while the Thouless pump assumes an infinite open coordinate space, the phase variable is confined to a periodic closed space. We show the counterintuitive result that quantum effects break the winding-number quantization, underscoring the fundamental difference between open- and closed-space Thouless pumps. In addition, we analyze non-adiabatic effects on phase dynamics and the role of Landau–Zener transitions in modifying winding numbers. Finally, we discuss the role of PT-symmetry in this quantum system in describing hysteretic time evolution.

## II. CLASSICAL TIME EVOLUTION OF "PHASE PARTICLE" IN GENERALIZED ADLER EQUATIONS

We start from the classical phase dynamics described by a generalized Adler equation. The equation is given by

$$\frac{d\varphi}{dt} = f(\varphi) = -\frac{\partial V(\varphi)}{\partial \varphi}, \qquad V(\varphi) = -\varphi\Delta + v(\varphi). \tag{1}$$

Here, $\varphi$ is a phase difference between two self-sustained oscillators, $f(\varphi) = f(\varphi + 2\pi)$ is a periodic force field, and $V(\varphi)$ is the potential with its periodic part, $v(\varphi) = v(\varphi + 2\pi)$, and $\Delta$ is a detuning between two oscillators. The dynamics corresponds to the overdamped motion of a "phase particle" driven by the generalized "washboard" potential $V(\varphi)$. The conventional Adler equation is given by choosing $v(\varphi) = A\cos\varphi$. By integrating this equation, we obtain

$$t_f - t_i = \int_{\varphi_i}^{\varphi_f} \frac{d\varphi}{f(\varphi)} \tag{2}$$

The integral logarithmically diverges at the point $\varphi = \varphi_0$, at which the force field vanishes: $f(\varphi_0) = 0$. This means that an infinitely long time is required for $\varphi$ to approach $\varphi_0$. That satisfying $f(\varphi_0) = 0$ and $\partial_\varphi f(\varphi_0) < 0$ (i.e., the local minima of the phase potential $V(\varphi)$) is an attractive fixed point, to which $\varphi$ asymptotically approaches and the two oscillators are synchronized [29]. In contrast, $\varphi$ departs from a repulsive fixed point, $f(\varphi_0) = 0$ and $\partial_\varphi f(\varphi_0) > 0$, so that the oscillators are not synchronized at the local maximum but only at the local minimum of $V(\varphi)$ [30].

When the force field has sufficiently slow time dependence, we can employ adiabatic approximation, where the phase trajectory is determined by tracing the instantaneous attractive fixed points, given by $f(\varphi, t) = 0$ with $\partial_\varphi f(\varphi) < 0$ (solid line in Fig.1a.). As suggested by M. V. Berry in 1985 [31], the phase progress of periodic oscillation in an adiabatically modified classical Hamiltonian system can also be divided into two contributions as in the case of a quantum system; one is the phase originated from its eigen frequency and the other is the additional one, called Hannay's angle [32]. The latter is regarded as the classical counterpart of the geometrical phase in quantum systems [31]. The time evolution of phase difference per modulation cycle discussed here can be analogous to Hannay's angle. It reflects the geometrical features in this driven-dissipative nonlinear system.

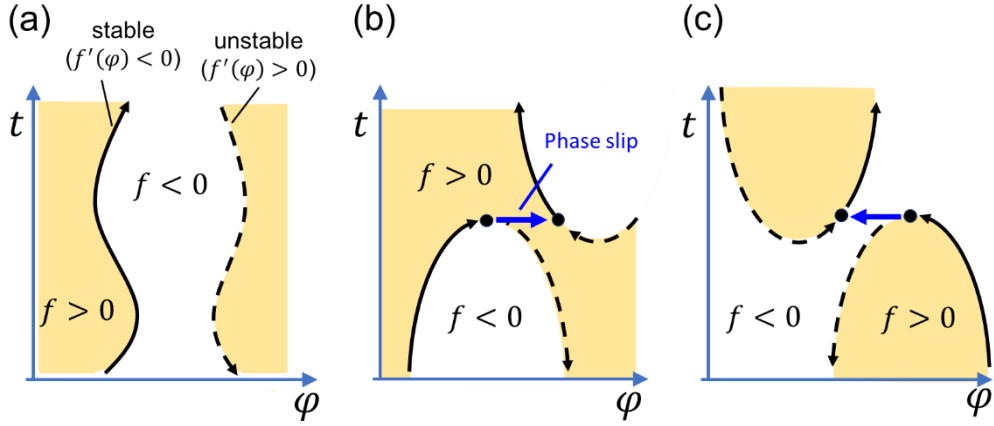

Fig.1 Schematic drawings of adiabatic trajectories of a "phase particle" under the time-dependent force field $f(\varphi,t)$. The solid (dashed) lines show the stable (unstable) trajectories for which $\partial_\varphi f(\varphi,\theta) < 0$ ($> 0$). (a) A stable adiabatic time evolution of the particle is induced when the force field has no singularity as a function of time. (b) and (c) Saddle-node bifurcation of fixed points induces the phase slip and its direction depends on the relative geometry of two domains, induces the positive (b) and negative (c) slips.

When the curve $f(\varphi,t) = 0$ is closed, i.e., showing a saddle-node bifurcation, the trajectory can no longer be stabilized and $\varphi$ slips to find another equilibrium point as shown in Fig.1(b) and (c), inducing the breakdown of adiabatic approximation. The slip direction is determined by the topological configuration of two domains, $f < 0$ and $f > 0$. When the latter (former) encloses the former (latter) around the saddle point as shown in Fig.1b (Fig.1c), the phase slips in the positive (negative) direction. If we expand $f(\varphi,t)$ at around the saddle point $(t_0, \varphi_0)$ as $f(\varphi,t) = a(t-t_0)(\varphi-\varphi_0)^2 + \cdots$, positive and negative slips are induced for $a > 0$ and $< 0$, respectively. The slip direction is robust against the small perturbation to the parameters, $a$, $t_0$, and $\varphi_0$, except for the point, $a \sim 0$, where a significant change emerges in the dynamics [30]. The response is analogous to topological phase transition in similar driven-dissipative nonlinear systems [33,34].

### III.   CLASSICAL TWO-TONE PUMPED ADLER EQUATION

We then investigate the detailed dynamics by using a specific example, i.e., an overtone-synthesized Adler equation [20],

$$\frac{d\varphi}{dt} = -\frac{\partial V(\varphi)}{\partial \varphi}, \quad V(\varphi, \theta(t)) = -\varphi\Delta - A_1 \cos(\varphi - \theta(t)) - A_2 \cos 2\varphi. \quad (3)$$

$\varphi$ is the phase difference between two self-sustained oscillations and $\Delta$ is their frequency detuning. $A_1$ and $A_2$ are amplitudes of fundamental and overtone couplings, respectively. $V(\varphi, \theta(t))$ has a time modulation through the normalized time $\theta(t) = \Omega t$, where $\Omega$ is the angular frequency of temporal modulation and is assumed to be small enough to maintain the adiabatic process. This generalized Adler equation well describes the phase synchronization and phase-slip dynamics experimentally observed in coupled optomechanical oscillators [20]. The detailed derivation of (3) starting from a parametrically coupled Duffing-Van der Pol equation is described in Section A in Supplemental Materials.

As already discussed, the steady-state trajectory of this "phase particle" can be obtained by finding the solution $\varphi(t)$ for the equation $f(\varphi(t), \theta(t)) = 0$, with $\partial_\varphi f(\varphi(t), \theta(t)) < 0$, except for the points of saddle-node bifurcation. Because the system describes a particle in a spatially and temporally periodic potential in the adiabatic limit, it can be regarded as a kind of classical analogy of the Thouless pump. The essential difference is that the phase space is finite. Although the Thouless pump also assumes a spatially periodic potential, the coordinate variable spans an infinite open space, which implies Bloch theorem to enable the continuous wave number to be defined. We discuss this point later in quantum dynamics.

The simplest cases are given by $A_2 = 0$ and $\Omega = 0$, corresponding to a conventional Adler equation. The synchronization condition is given by $-\partial_\varphi V(\varphi) = \Delta + A_1 \sin \varphi = 0$, which has the solution $\varphi = -\sin^{-1}(\Delta/A_1)$ within the area of "Arnold tongue", $|\Delta/A_1| \leq 1$. For a finite but sufficiently small modulation, $\Omega \ll A_1$, the phase synchronization is instantaneously satisfied, and the stabilized phase difference adiabatically changes with time as discussed in the previous section.

Figure 2 summarizes the trajectories calculated by numerically solving (3) as a function of $\varphi$ and the normalized time $\theta$. We assumed no detuning ($\Delta = 0$) for simplicity. The stabilized phase with the potential $V(\varphi)$ [(a) and (d)] and the trajectories with the landscape of force fields $f(\varphi, \theta)$ [(b) and (e)] are imaged. The same trajectories are also mapped on the $\theta - \varphi$ torus surface in Fig. 2(c) and (f). The two-tone mixing ratio $r$ is defined by $A_1 = \mu r$ and $A_2 = \mu(1-r)/2$, where the parameter $\mu$ is fixed to be unity in the calculation without loss of generality [35]. The force field is then given by $f(\varphi, \theta) = -[r \sin(\varphi - \theta) + (1-r)\sin(2\varphi)]$. As already discussed in the previous section, a significant change in the trajectory is observed at $r = r_c = 0.5$. When $r < r_c$ (Fig. 2a. – c.), the curve $f(\varphi, \theta) = 0$ does not show saddle-node bifurcation or winding in the $\varphi$ direction. In contrast, for $r > r_c$ (Fig. 2d. – f.), the curve $f(\varphi, \theta) = 0$ shows a bifurcation, leading to a discontinuous phase slip to the next steady-state phase (light thick blue arrow in Fig. 1d). The phase slip is visible also in Fig. 2(e) and (f) as the trajectories parallel to $\varphi$ axis. The discontinuous transition shows hysteretic behavior as shown by the difference between red (forward sweep: $\theta(t) = \Omega t$ with $\Omega > 0$) and blue (backward sweep: $\theta(t) = -\Omega t$) trajectories. The winding number, defined as $\chi \coloneqq (2\pi)^{-1} \int_{t_0}^{t_0+\frac{2\pi}{\Omega}} f(\varphi(t)) dt = (2\pi)^{-1} \int_{\varphi(t_0)}^{\varphi\left(t_0+\frac{2\pi}{\Omega}\right)} d\varphi$, is quantized to 0 or $\pm 1$ and shows discontinuous transition at $r_c$ [Fig. 3(a) and dotted line in Fig. 3(c)]. These results well reproduce reported experimental results [20], reflecting the geometrical characteristics of the force field landscape.

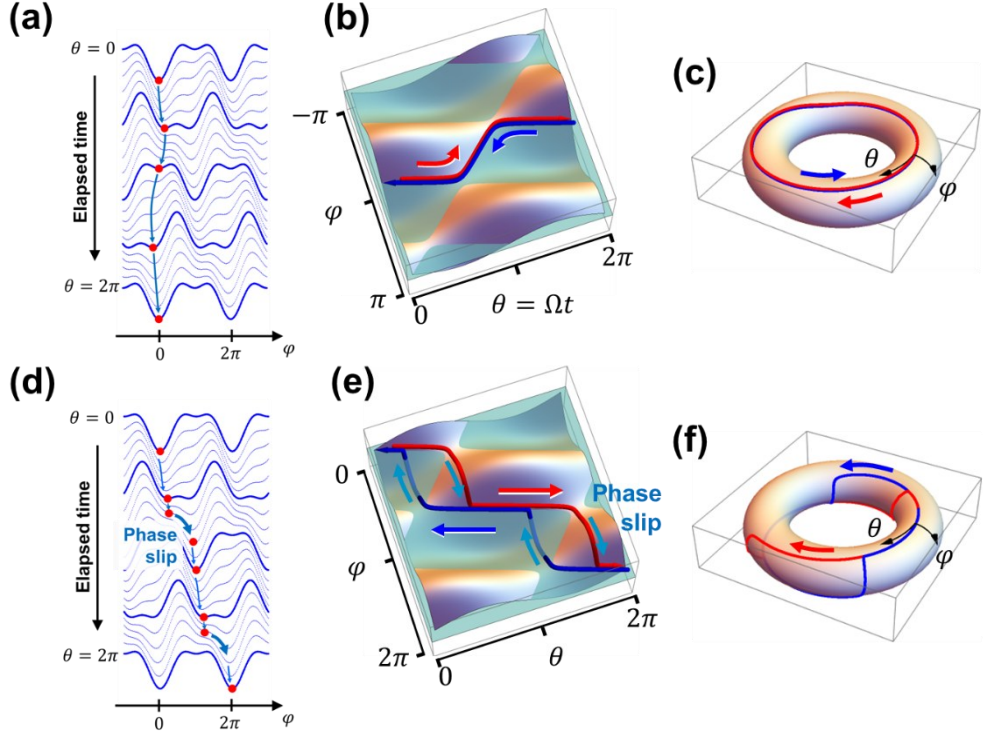

Fig.2 (a) and (d): The time evolution of $\varphi$ (red marks) calculated by numerically solving the Adler equation with generalized potentials (3) (blue curves). (b) and (e) The calculated trajectories superimposed on a bird's-eye view of the force field $f(\varphi, \theta)$. A horizontal plane ($f(\varphi, \theta) = 0$), by which the negative region of the force field is colored partially transparent light green, is also shown. The trajectory calculated for the reversed motion, i.e., $\theta(t) = -\Omega t$, is also shown as blue curves in addition to the forward motion, $\theta(t) = \Omega t$, colored red. (c) and (f) The trajectories are mapped on $\theta - \varphi$ torus surfaces. The used parameters are $\Delta = 0$, $r = 0.49$, (a, b, c) and $r = 0.51$ (d, e, f).

A nonzero detuning $\Delta \neq 0$ applies an additional constant force to the "phase particle", and the equation describes its motion in a tilted washboard potential [20]. The temporal modulation together with the detuning pumps the unidirectional motion of the "phase particle" even within the synchronization region of Arnold tongue, and clear non-reciprocity is observed depending on the sign of $\Delta$ (See Section B in Supplemental Materials). This non-reciprocal phase pumping corresponds to the motion in a "ratchet" potential and has also been experimentally observed in optomechanical experiments [20].

## IV. QUANTUM PUMP IN ADIABATIC LIMIT

As described in the previous section, the closed-space periodic pump quantizes the winding number in the classical limit. This is because the trajectory of a "phase particle" is deterministically defined by classical dynamics, and the temporal periodicity forces the steady-state adiabatic trajectory to be identical in every period. However, in a quantum system, the trajectory has a probability distribution, and the linear combination of different particle flows can also be a Floquet eigenstate. Therefore, the average winding number per cycle can be a non-integer even though the time evolution of the wave function is identically repeated in every period. This was first suggested in 1983 by Thouless, who presented a simple example with a perturbation calculation using a delta-function-shaped potential [21].

To theoretically describe the synchronization in the quantum domain, a standard method is to start with the original oscillator variables, i.e., the quadrature operators, and then study the quantum dynamics by using the quadrature Hamiltonian [36-39]. However, it is not straightforward to clarify the connection to the framework of an open-space Thouless pump because the spatial position of a particle corresponds to the phase in our closed synchronization system. Therefore, we employ a simple phase-only description while disregarding the amplitude, which was introduced also to study the synchronization in a semiclassical Kuramoto model [40].

Following the general formulation of the Thouless pump [21,22], we employ the quantum adiabatic approximation. The time-dependent Hamiltonian corresponding to the classical equation (3) is given by

$$\hat{H}(t) = \frac{\hat{p}_\varphi^2}{2m_e} - \mu \left[ r \cos(\hat{\varphi} - \Omega t) + \frac{1-r}{2} \cos 2\hat{\varphi} \right]. \tag{4}$$

Here, the "hat" on the variables is used to specify operators. Because we cannot directly define the Hermitian phase operator $\hat{\varphi}$ that satisfies the periodic boundary condition [41], we only introduce trigonometric operators, $e^{\pm i\hat{\varphi}}$, as well as $\cos\hat{\varphi}$, and $\sin\hat{\varphi}$, and also the canonical conjugate of $\hat{\varphi}$ as $\hat{p}_\varphi = -i\partial_\varphi$, which satisfy the commutation relations, $[\hat{p}_\varphi, e^{\pm i\hat{\varphi}}] = \pm e^{\pm i\hat{\varphi}}$. The detailed formulation of phase operators starting from a tight-binding model in reciprocal lattice space is described in Section C in Supplemental Materials. In this quantum model, $m_e$, the mass of the "phase particle", is newly introduced, because the wave packet broadening needs to be parametrized. However, the results of adiabatic approximation are not altered if we keep $m_e\mu$ constant. Therefore, we again fix $\mu = 1$ and examine the effects of wave packet broadening by simply changing $m_e$. Reducing $m_e$ causes the wave function enhancing the quantum effects to broaden. It is also important to satisfy the periodic boundary condition in the definition of the velocity operator, $\hat{v}_\varphi$, which plays an essential role in deriving the current flow. We can define it as $\hat{v}_\varphi = \hat{p}_\varphi/m_e$, which is consistent with Heisenberg's equation of motion for the phase operators (Section C in Supplemental Materials).

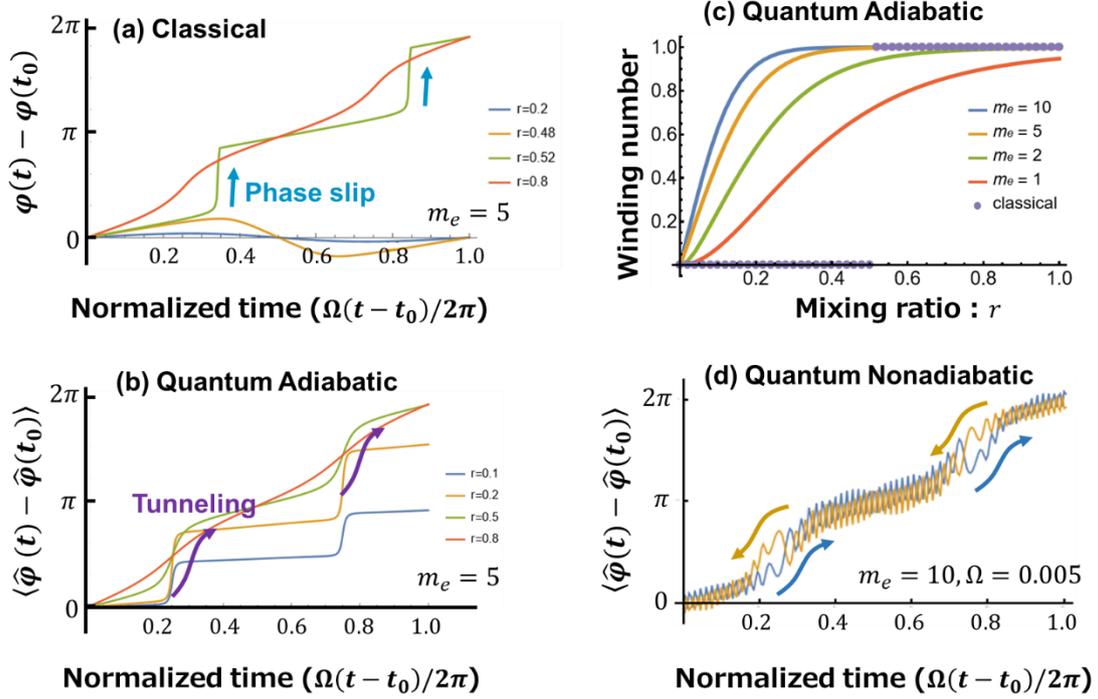

Fig. 3. (a) Calculated phase evolution using the classical model (3) with various $r$. The winding number, $\chi = [\varphi(t_0 + 2\pi\Omega^{-1}) - \varphi(t_0)]/2\pi$, shows a discontinuous transition at $r = r_c = 0.5$. (b) Calculated phase evolution under quantum adiabatic approximation for Hamiltonian (4) with various $r$ with $m_e = 5$. (c) Detailed $r$ dependence of average phase shift per cycle, i.e., the winding number, for various $m_e$. That calculated by classical model (3) is also shown by the dotted plot for comparison. (d) The phase evolution by non-adiabatic Floquet theory for forward (blue: $\theta(t) = \Omega t$) and backward (yellow: $\theta(t) = -\Omega t$) sweeps with $m_e = 10$ and $\Omega = 0.005$. The Landau-Zener-Stückelberg interference is visible in the hysteresis region.

We then apply a first order correction to the time-dependent state vector in adiabatic approximation, as usually employed in the formulation of the Thouless pump [21]. Introducing the instantaneous $n$-th order eigenstate and eigen energy of $\hat{H}(t)$ as $|\phi_n(t)\rangle$ and $E_n(t)$, satisfying $\hat{H}(t)|\phi_n(t)\rangle = E_n(t)|\phi_n(t)\rangle$, the first order correction to the ground state is given by

$$|\psi_0^{(1)}(t)\rangle = |\phi_0(t)\rangle - i \sum_{m \neq 0} |\phi_m(t)\rangle \frac{\langle \phi_m(t)|\vec{\partial_t}|\phi_0(t)\rangle}{E_0(t) - E_m(t)}, \quad (5)$$

and the lowest-order velocity can be calculated as the expectation value of $\hat{v}_\varphi = \hat{p}_\varphi/m_e$ as

$$\langle \psi_0^{(1)}(t)|\hat{v}_\varphi|\psi_0^{(1)}(t)\rangle \sim -i \frac{1}{m_e} \sum_{m \neq 0} \langle \phi_0(t)|\hat{p}_\varphi|\phi_m(t)\rangle \frac{\langle \phi_m(t)|\vec{\partial_t}|\phi_0(t)\rangle}{E_0(t) - E_m(t)} + h.c. . \quad (6)$$

The phase evolution as a function of elapsed time can be calculated by integrating (6) with time. Figure

3(b) shows the example of calculated phase evolution, i.e., the integration of (6) from $t_0 = 0$, for the Hamiltonian (4) with various $r$. The detailed $r$ dependence of phase shift per cycle is shown in Fig. 3(c). In contrast to the classical case, the winding number $\chi := \int_{t_0}^{t_0+\frac{2\pi}{\Omega}} \langle \psi_0^{(1)}(t) | \hat{v}_\varphi | \psi_0^{(1)}(t) \rangle dt / 2\pi$ is *not* quantized to be an integer, and its $r$ dependence is not discontinuous. The broken quantization of the winding number in the quantum system sounds counterintuitive, but it can be understood by considering the superposition of different winding-number states. Figure 4 shows the mapping of wave function amplitude in 2D $\varphi - \theta$ space and their snapshots slightly before and after the slip angle, $\theta = 3\pi/2$. Two cases are compared: $m_e = 10$ and $r = 0.55$ leading to a winding number of 1.00 [(a) and (b)], and $m_e = 1$ and $r = 0.55$, to a non-integer winding number of 0.76 [(c) and (d)]. In the former, the wave function is well localized, as shown by the solid lines in (b) (here the purple dashed $E_0$-line shows the zero-amplitude axis), and the direction of pump-induced rotation is well defined. On the other hand, the latter shows the wave function to extend over the whole phase space (the solid lines in (d)). With these parameters, the contributions from different winding numbers are superimposed, leading to the "unquantized" winding number. The feature of a non-integer winding number is the consequence of the quantum mechanical broadening of wave packets.

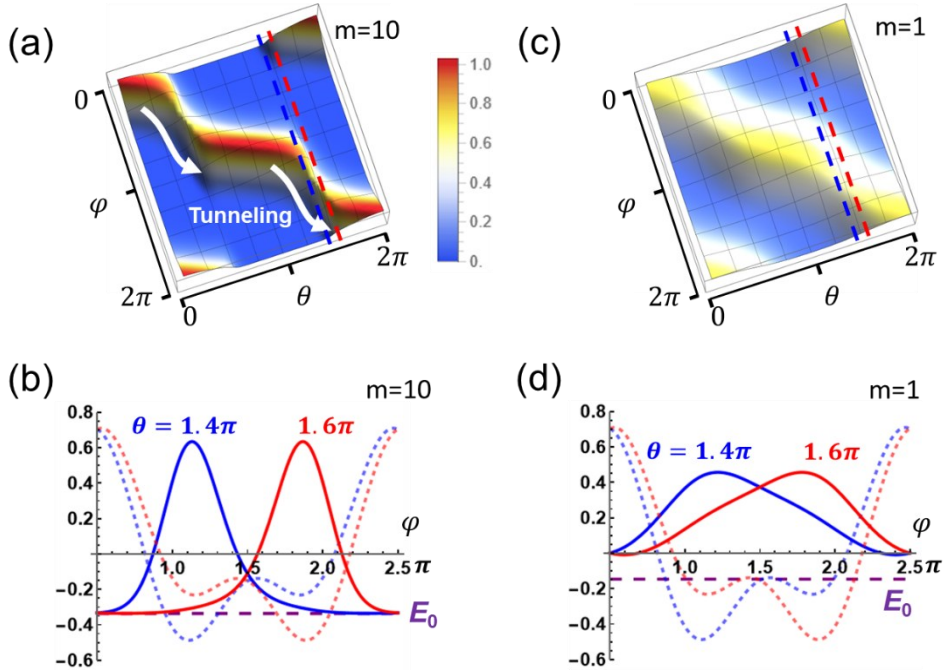

Fig. 4. (a) and (c): The calculated time evolution of the amplitude of instantaneous wave function using an adiabatic approximation with $m_e = 10$ (a) and 1 (c) with $r = 0.55$. (b) and (d): The amplitude of wave function at specific values of $\theta$ (1.4$\pi$ and 1.6$\pi$), which correspond to the colored dashed lines in (a) and (c). The instantaneous eigen energy $E_0$ (purple dashed lines) relative to the potential $V(\varphi)$ (red and blue dashed lines) are also shown. The amplitude curves are shifted

downward to align $E_0$ lines to their horizontal axis.

## V. NON-ADIABATIC EFFECTS AND PT-SYMMETRY

The adiabatic approximation assumes perfect tunneling even if a finite energy-barrier is between two local minimums in the potential. In this approximation, the phase is never kept trapped by one of the two minimums in the double well potential and the winding number cannot vanish, even for small $r$ and large $m_e$, inducing a large difference from the classical case even when the wave function is highly localized. However, with a finite modulation frequency, the coherent oscillation between two local minimums, i.e., the Landau-Zener-Stückelberg interference, allows the backward tunneling to modify the overall winding number. We investigate this non-adiabatic effect by directly applying the Floquet theory [42-43] to our overtone-synthesized Adler model to solve the time-dependent Schrödinger equation in 2D $\varphi - \theta$ space.

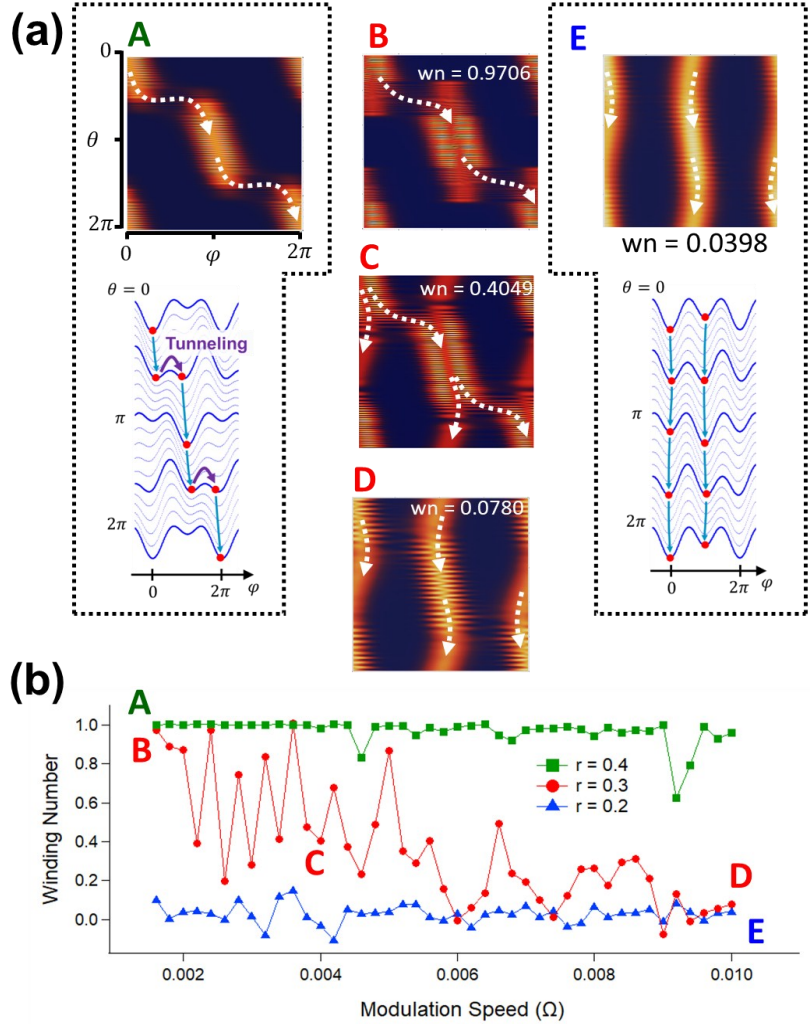

Fig. 5. (a) Time evolution of wave function calculated with various values of $r$ and $\Omega$ with $m_e = 10$. "wn" shown in the plots quantifies the calculated winding number. Labels A - E correspond to the parameters shown in (b). Also shown is the schematic drawing of the motion of a phase particle in the potential shape for the parameters A and E. (b) Detailed plots of calculated winding numbers as a function of modulation speed with $m_e = 10$.

The details of the calculation procedure are described in Section D in Supplemental Materials. Figure 5(a) shows a two-dimensional map of the amplitude square of the wave function calculated for various parameters, with schematic illustrations of the phase flow with respect to the potential. Figure 5b. shows the calculated winding number for a range of modulation speed $\Omega$. When the temporal modulation is sufficiently strong and slow (for example $r = 0.4$ and $\Omega = 0.0016$ shown as A in Fig. 5), the time evolution of the wave function reproduces the results by quantum adiabatic approximation, and a unit winding number is obtained. Here, the quantum tunneling induces a phase slip similarly to the classical system. On the other hand, with a reduced modulation depth (for example $r = 0.2$ shown as E in Fig. 5a.), tunneling probability between two potential minimums becomes negligibly small, leading to a small winding number. This is contrary to the results of adiabatic approximation shown in Fig. 2c, where a nearly-unity winding number is always obtained, but approaching the classical limit. In the area between them (for example, $r = 0.3$ shown as B-D in Fig. 5a.), the winding number rapidly changes as a function of $\Omega$. This should be compared to the case of an infinite open-space Thouless pump [44], where the non-adiabatic effect only slightly modifies the winding number. In the case of an open-space pump, the contributions from the wave number in the vicinity of the Brillouin Zone compensate for the reduction of current density at gamma point. In our closed-space Thouless pump, the continuous transition from unity to a zero winding number rapidly occurs because only the gamma point ($k = 0$) contribution governs the total current flow.

Finally, we comment on the role of *PT*-symmetry in this closed-space Thouless pump. The Hamiltonian (4) is *PT*-symmetric and invariant under the simultaneous reversal of $\hat{\varphi} \to -\hat{\varphi}$ and $t \to -t$. (Note that the concept is different from generally discussed non-Hermitian quantum systems, where the gain and loss are balanced to maintain the symmetry, for example see [45]). This implies the quasi-energy eigen function, $\psi_n(\varphi, t) := \langle \varphi | \psi_n(t) \rangle$, to be the eigen function of *PT* reversal operation, $\mathcal{PT}\psi_n(\varphi, t) := [\psi_n(-\varphi, -t)]^* = \lambda \psi_n(\varphi, t)$, where the eigen value of reversal operation, $\lambda$, has a unit magnitude and is phase dependent (Section E in Supplemental Materials). Our calculation in fact numerically reproduces the *PT* invariance for all the quasi-energy eigen functions. The hysteretic behavior observed in the classical limit [20] looks inconsistent with this consequence, but the linear superposition of two quasi-energy eigenstates well reproduces the behavior (Fig. 3(d)). This superposition state is not the eigenstate of $\mathcal{PT}$ operation, being consistent with the non-reciprocal hysteretic behaviors. The Landau-Zener-Stückelberg interference is visible in this hysteresis region, showing the feature of quasi-energy non-eigenstates. The relation between the gauge transformation and non-reciprocal hysteretic time evolution under *PT*-symmetric system is an interesting subject but should be discussed in detail elsewhere.

## VI. CONCLUSION

We studied classical and quantum phase dynamics described by a generalized Adler equation. In the classical regime, the winding number of its deterministic trajectory is quantized to be an integer, showing discontinuous transition as a function of system parameters. The transition is caused by the change in the topology of the geometrical landscape of a force field mapped in a 2D $\varphi - \theta$ space, leading to the non-adiabatic phase slip induced by the saddle-node bifurcation. The quantization is broken in the quantum regime, because the quasi-energy eigenstates become the superposition of different winding number states. The hysteric behavior at the phase slip is reproduced in non-adiabatic calculations, where the Landau-Zener-Stückelberg oscillation replaces the deterministic phase slip observed in the classical systems.

## ACKNOWLEDGEMENTS

The authors thank Hajime Okamoto for his continuous encouragement. This work was partially supported by JSPS KAKENHI Grant Number JP21H05020 and JP23H05463.

# REFERENCES

1. J. H. Vincent, "On some experiments in which two neighboring maintained oscillatory circuits affect a resonating circuit", Proc. Phys. Soc. London 32, 84 (1919).
2. R. Adler, Proc. "A study of locking phenomena in oscillators", Proc. IRE 34, 351 (1946) [Proc. IEEE 61, 1380 (1973)].
3. H. L. Stover and W. H. Steier, "Locking of laser oscillators by light injection", Appl. Phys. Lett. 8, 91 (1966).
4. K. Kurosawa, "Injection locking of microwave solid-state oscillators", Proc. IEEE 61, 1386 (1973).
5. H. W. Yen and M. K. Barnoski, "Optical injection locking and switching of transistor oscillators", Appl. Phys. Lett. 32, 182 (1978).
6. M. Zalalutdinov et al., "Frequency entrainment for micromechanical oscillator", Appl. Phys. Lett. 83, 3281 (2003).
7. W. H. Rippard et al., "Injection Locking and Phase Control of Spin Transfer Nano-oscillators", Phys. Rev. Lett. 95, 067203 (2005).
8. S. Kaka et al., "Mutual phase-locking of microwave spin torque nano-oscillators", Nature 437, 389 (2005).
9. S. B. Shim, M. Imboden, P. Mohanty, "Synchronized Oscillation in Coupled Nanomechanical Oscillators", Science 316, 95 (2007).
10. M. Zhang, G. S. Wiederhecker, S. Manipatruni1, A. Barnard, P. McEuen, and M. Lipson, "Synchronization of Micromechanical Oscillators Using Light", Phys. Rev. Lett. 109, 233906 (2012).
11. M. J. Seitner et al., "Parametric oscillation, frequency mixing, and injection locking of strongly coupled nanomechanical resonator modes", Phys. Rev. Lett. 118, 254301 (2017).
12. Y. Kuramoto, "Cooperative dynamics of oscillator community: a study based on lattice of rings", Prog. Theor. Phys. Suppl. 79, 223 (1984).
13. A. S. Pikovsky, M. Rosenblum, and J. Kurths, *Synchronization: A Universal Concept in Nonlinear Science* (Cambridge University Press, New York, 2001).
14. F. A. Rodrigues, T. K. DM. Peron, Peng Ji, J. Kurths "The Kuramoto model in complex networks", Phys. Rep. 610, 1 (2016).
15. D. Cumin, C.P. Unsworth, "Generalising the Kuramoto model for the study of neuronal synchronization in the brain", Physica D 226, 181 (2007).
16. M. Chen, "Synchronization in time-varying networks: A matrix measure approach", Phys. Rev. E 76, 016104 (2007).
17. S. Petkoski and A. Stefanovska, "Kuramoto model with time-varying parameters", Phys. Rev. E 86, 046212 (2012).

18. S. Urazhdin, P. Tabor, V. Tiberkevich, and A. Slavin, "Fractional synchronization of spin-torque nano-oscillators", Phys. Rev. Lett. 105, 104101 (2010).

19. M. H Matheny et al., "Exotic states in a simple network of nanoelectromechanical oscillators" Science 363, eaav7932 (2019).

20. M. Asano, H. Okamoto, and H. Yamaguchi, "Synthesized Kuramoto potential via optomechanical Floquet engineering", Science Adv. 11, ady4167 (2025).

21. D. J. Thouless, "Quantization of particle transport", Phys. Rev. B 27, 6083 (1983).

22. R. Citro and M. Aidelsburger, "Thouless pumping and topology", Nature Review Phys. 5, 87–101 (2023).

23. M. Lohse, C. Schweizer, O. Zilberberg, M. Aidelsburger, and I. Bloch, "A Thouless quantum pump with ultracold bosonic atoms in an optical superlattice", Nature Phys. 12, 350–354 (2016).

24. S. Nakajima, T. Tomita, S. Taie, T. Ichinose, H. Ozawa, L. Wang, M. Troyer, and Y. Takahashi, "Topological Thouless pumping of ultracold fermions", Nature Phys. 12, 296–300 (2016).

25. O. Zilberberg et al., "Photonic topological boundary pumping as a probe of 4D quantum Hall physics", Nature 553, 59 (2018).

26. Z. Fedorova, H. Qiu, S. Linden, and J. Kroha, "Observation of topological transport quantization by dissipation in fast Thouless pumps", Nature Commun. 11:3758 (2020).

27. A. Cerjan, M. Wang, S. Huang, K. P. Chen, and M. C. Rechtsman "Thouless pumping in disordered photonic systems", Light: Science & Applications, 9:178 (2020).

28. O. You, S. Liang, B. Xie, W. Gao, W. Ye, J. Zhu, and S. Zhang, "Observation of Non-Abelian Thouless Pump", Phys. Rev. Lett. 128, 244302 (2022).

29. F. Dörfler and F. Bullo, "Synchronization in complex networks of phase oscillators: A survey", Automatica 50, 1539–1564 (2014).

30. S. Wiggins, *Introduction to Applied Nonlinear Dynamical Systems and Chaos*, 2nd Eds. Sec.20.1 (Springer-Verlag, New York 2003).

31. M. V. Berry, "Classical adiabatic angles and quantal adiabatic phase", J. Phys. A: Math. Gen. 18, 15 (1985).

32. J. H. Hannay, "Angle variable holonomy in adiabatic excursion of an integrable Hamiltonian", J. Phys. A: Math. Gen. 18, 221 (1985).

33. V. Dumont, M. Bestler, L. Catalini, G. Margiani, O. Zilberberg, and A. Eichler, "Energy landscape and flow dynamics measurements of driven-dissipative systems" Phys. Rev. Res. 6, 043012 (2024).

34. G. Villa, J. del Pino, V. Dumont, G. Rastelli, M. Michałek, A. Eichler, O. Zilberberg, Sci. Adv. "Topological classification of driven-dissipative nonlinear systems", Sci. Adv. 11, eadt9311 (2025).

35. This is because its ratio to $\Omega$ changes the relaxation rate but does not change the trajectories in

$\varphi - \Omega t$ space in adiabatic limit ($\Omega/k \ll 1$).

36. S. Walter, A. Nunnenkamp, and C. Bruder, "Quantum Synchronization of a Driven Self-Sustained Oscillator" Phys. Rev. Lett. 112, 094102 (2014).
37. E. Amitai, N. Lörch, A. Nunnenkamp, S. Walter, and C. Bruder, "Synchronization of an optomechanical system to an external drive", Phys. Rev. A 95, 053858 (2017).
38. A. Roulet and C. Bruder, "Quantum Synchronization and Entanglement Generation", Phys. Rev. Lett. 121, 063601 (2018).
39. O.V. Zhirov and D.L. Shepelyansky, "Quantum synchronization", Eur. Phys. J. D 38, 375–379 (2006).
40. I. H. de Mendoza, L. A. Pachón, J. Gómez-Gardeñes, and D. Zueco, "Synchronization in a semiclassical Kuramoto model", Phys. Rev. E 90, 052904 (2014).
41. L. Susskind and J. Glogower, "Quantum mechanical phase and time operator", Physics 1, 49-61 (1964).
42. H. Sambe, "Steady states and quasienergies of a quantum-mechanical system in an oscillating field" Phys. Rev. A 7, 2203 (1973).
43. A. Eckardt, "Colloquium: Atomic quantum gases in periodically driven optical lattices", Rev. Mod. Phys. 89, 011004 (2017).
44. L. Privitera, A. Russomanno, R. Citro, and G. E. Santoro "Nonadiabatic Breaking of Topological Pumping", Phys. Rev. Lett. 120, 106601 (2018).
45. V. V. Konotop, J. Yang, and D. A. Zezyulin, "Nonlinear waves in PT-symmetric systems", Rev. Mod. Phys. 88, 035002 (2016).

# Supplemental Materials

## A. Derivation of Adler equation from parametrically coupled self-sustained oscillation

We here derive the generalized Adler equation from the model of parametrically coupled self-sustained oscillators by extending the formulation by R. Lifshits, E. Kenig, and M. C. Cross [S1].

1. Single Duffing-Van der Pol oscillator

We start from the rotating-wave approximation to a single Duffing-Van der Pol oscillator with negative and nonlinear damping:

$$\ddot{u} + \omega_0^2 u - \gamma \omega_0 \dot{u} + b\omega_0 u^2 \dot{u} + a\omega_0^2 u^3 = 0. \tag{S1}$$

Here, $u(t)$ is the oscillator dynamical variable and $\gamma > 0$ is the dimensionless damping. $a$ and $b$ are dispersive and dissipative nonlinear coefficients with the dimension of $[u]^{-2}$. The negative damping in the third term represents positive feedback inducing a self-sustained oscillation, whereas the nonlinear damping in the fourth term saturates the steady-state amplitude. By redefining, $u \to \sqrt{\gamma/b} \cdot u$ and $a \to ab/\gamma$, both $a$ and $u$ become dimensionless, and we obtain

$$\ddot{u} + \omega_0^2 u - \gamma \omega_0 (1 - u^2)\dot{u} + a\omega_0^2 u^3 = 0. \tag{S2}$$

Here, we assume that $a \ll 1$ and $\gamma \ll 1$. Starting from a small vibration amplitude, the negative damping increases the amplitude until the saturation is induced by the third term, where $\overline{u^2} \sim 1$. Therefore, the steady-state vibrational amplitude is in the order of $O(1)$ if $a$ is sufficiently small. By using a sinusoidal solution, $u \sim u_0 e^{i\omega t} + c.c. = Re[2u_0 e^{i\omega t}]$, (S2) becomes

$$(\omega_0^2 - \omega^2)u_0 + i\omega \omega_0 \gamma(|u_0|^2 - 1)u_0 + 3a\omega_0^2 |u_0|^2 u_0 = 0. \tag{S3}$$

Thus, the steady-state solution is given by

$$|u_0| = 1, \quad \omega = \omega_0\sqrt{1 + 3a} \sim \omega_0\left(1 + \frac{3a}{2}\right). \tag{S4}$$

Then we consider the case that the oscillator is nearly steady-state but the amplitude and phase slowly change with time:

$$u(t) \sim [u_0(t) e^{i\omega_0 t} + c.c.]. \tag{S5}$$

This rotating frame approximation assumes that the time variation of $u_0(t)$ is sufficiently slower than $\omega_0$.

$$|\dot{u}_0(t)| \ll \omega_0 |u_0(t)|. \tag{S6}$$

We introduce a new scaling variable $\varepsilon$ and a scaled dimensionless time $T = \varepsilon \omega_0 t$, and $U(T) \coloneqq u_0(t) = u_0(T/\varepsilon\omega_0)$. Then

$$\frac{du_0(t)}{dt} = \frac{dT}{dt}\frac{dU(T)}{dT} = \varepsilon\omega_0 \frac{dU(T)}{dT} := \varepsilon\omega_0 U'(T). \tag{S7}$$

Here, $\varepsilon$ can be chosen in such a way that $|U'(T)/U(T)|$ becomes less than the order of unity. This means that $\varepsilon$ is chosen to be small, i.e., $\varepsilon \ll 1$, as assumed by (S6). With these variables, the rotating-wave approximation is given by neglecting second and higher order terms with respect to $\varepsilon$. In addition, we also neglect the terms $\varepsilon\gamma$ and $\varepsilon a$ and their higher order terms because we consider the cases $a \ll 1$ and $\gamma \ll 1$. We then extend (S2) to the case with an additional detuning $\Delta$,

$$\ddot{u} + \omega_0^2(1+\Delta)u - \gamma\omega_0(1-u^2)\dot{u} + a\omega_0^2 u^3 = 0. \tag{S8}$$

By using (S5)-(S8) and the definition of $U(T)$, we approximately obtain

$$2i\varepsilon U' + \Delta U + i\gamma(|U|^2 - 1)U + 3a|U|^2 U = 0. \tag{S9}$$

This equation leads to

$$\frac{U'}{U} = \frac{\gamma}{2\varepsilon}(1 - |U|^2) + i\left(\frac{\Delta}{2\varepsilon} + \frac{3a}{2\varepsilon}|U|^2\right). \tag{S10}$$

The second term in r.h.s. of (S10) induces the oscillation of $U$, and the first term does the damping or amplification depending on the sign of $1 - |U|^2$. Therefore, we can assume that $|U|$ has the order of unity in the steady state. To choose $\varepsilon$ in such a way that the order of $|U'/U|$ becomes less than unity, the order of three parameters,

$$D := \frac{\Delta}{\varepsilon}, \qquad \Gamma := \frac{\gamma}{\varepsilon}, \qquad \alpha := \frac{a}{\varepsilon}, \tag{S11}$$

must be less than unity. Therefore, we need to assume that the three parameters, $\Delta$, $\gamma$, $a$, are all sufficiently smaller than unity and $\varepsilon \ll 1$ has the same order as the largest one.

If we assume the harmonic oscillation, $U(T) \sim U_0 e^{i\Omega T}$,

$$-2\Omega + D + i\Gamma(|U_0|^2 - 1) + 3\alpha|U_0|^2 = 0. \tag{S12}$$

Both the real and imaginary parts should vanish then we obtain $U_0 = e^{i\delta}$, and $D - 2\Omega + 3\alpha = 0 \rightarrow \Omega = (3\alpha + D)/2$. They lead to

$$u_0(t) = U_0 e^{i\Omega T} = e^{i\delta} e^{i\frac{3\alpha+D}{2}\varepsilon\omega_0 t} = e^{i\left(\frac{3a+\Delta}{2}\omega_0 t + \delta\right)}. \tag{S13}$$

This is consistent with (S4) with an additional detuning $\Delta\omega_0/2$. (S10) or the reformed equation

$$iU' = -\frac{D}{2}U + i\frac{\Gamma}{2}(1 - |U|^2)U - \frac{3\alpha}{2}|U|^2 U. \tag{S14}$$

gives the basic equation for the Duffing-Van der Pau oscillator in rotating-wave approximation.

2. Coupled Duffing-Van der Pol oscillators

Next, we consider the case in which two oscillators are lineally coupled with a detuning $\Delta$.

$$\ddot{u}_1 + \omega_0{}^2(1-\Delta)u_1 - \gamma\omega_0\dot{u}_1 + b\omega_0 u_1{}^2\dot{u}_1 + a\omega_0{}^2 u_1{}^3 = \kappa\omega_0{}^2 u_2, \quad (S15)$$
$$\ddot{u}_2 + \omega_0{}^2(1+\Delta)u_2 - \gamma\omega_0\dot{u}_2 + b\omega_0 u_2{}^2\dot{u}_2 + a\omega_0{}^2 u_2{}^3 = \kappa\omega_0{}^2 u_1.$$

We assume the forms $u_1(t) \sim [U_1(T)e^{i\omega_0 t} + c.c.]$, $u_2(t) \sim [U_2(T)e^{i\omega_0 t} + c.c.]$, then the equation of motion in the rotating frame is

$$iU_1' = \frac{D}{2}U_1 + i\frac{\Gamma}{2}(1-|U_1|^2)U_1 - \frac{3\alpha}{2}|U_1|^2 U_1 + \frac{K}{2}U_2,$$
$$iU_2' = -\frac{D}{2}U_2 + i\frac{\Gamma}{2}(1-|U_2|^2)U_2 - \frac{3\alpha}{2}|U_2|^2 U_2 + \frac{K}{2}U_1. \quad (S16)$$

Here, $K \coloneqq \kappa/\varepsilon$. Decomposing the complex amplitude into the amplitude and phase, $U_n(T) = R_n(T)e^{i\varphi_n(T)}$ for $n = 1,2$, we obtain

$$iR_1' - \varphi_1' R_1 = \frac{D}{2}R_1 + i\frac{\Gamma}{2}(1-|R_1|^2)R_1 - \frac{3\alpha}{2}|R_1|^2 R_1 + \frac{K}{2}R_2 e^{i(\varphi_2-\varphi_1)},$$
$$iR_2' - \varphi_2' R_2 = -\frac{D}{2}R_2 + i\frac{\Gamma}{2}(1-|R_2|^2)R_2 - \frac{3\alpha}{2}|R_2|^2 R_2 + \frac{K}{2}R_1 e^{i(\varphi_1-\varphi_2)}. \quad (S17)$$

Decomposing the real and imaginary parts of the first equation,

$$-\varphi_1' = \frac{D}{2} - \frac{3\alpha}{2}|R_1|^2 + \frac{KR_2}{2R_1}\cos(\varphi_2 - \varphi_1),$$
$$1 - |R_1|^2 \sim \frac{KR_2}{\Gamma R_1}\sin(\varphi_2 - \varphi_1). \quad (S18)$$

Here, we assume that the amplitude is nearly saturated and $R_1'/R_1$ is negligibly small. From the similar formula for $\varphi_2$, we finally obtain

$$\frac{d(\varphi_2 - \varphi_1)}{dT} = D - \frac{3\alpha K}{2\Gamma}\left(\frac{R_1}{R_2} + \frac{R_2}{R_1}\right)\sin(\varphi_2 - \varphi_1) + \frac{K}{2}\left(\frac{R_2}{R_1} - \frac{R_1}{R_2}\right)\cos(\varphi_2 - \varphi_1) \quad (S19)$$

If we simply assume two identical oscillators, $R_1 \sim R_2$, the cosine term vanishes and

$$\frac{d\varphi}{dT} = D - C\sin\varphi. \quad (S20)$$

Here, $C \sim \frac{3AK}{\Gamma}$ and $\varphi = \varphi_2 - \varphi_1$. From the derivation, we can find that the phase synchronization term, $C\sin\varphi$, is not directly caused by the linear coupling but by the amplitude modulation, which is proportional to $\sin\varphi$, through the Duffing nonlinearity. This modulation provides the opposite contribution between $\varphi_1$ and $\varphi_2$, then the two phases are intended to be identical in their time evolution. Eq. (S20) can be expressed by using the original parameters as

$$\omega_0{}^{-1}\frac{d\varphi}{dt} = \Delta - \frac{3a\kappa}{\gamma}\sin\varphi, \quad (S21)$$

which gives the Adler equation.

3. Parametric coupling

Next, we consider the case of parametric coupling, where the two modes have different frequencies, $\omega_1$ and $\omega_2$. The coupling is mediated by oscillating linear coupling constants, given by

$$k_1(t) = \kappa_1 e^{i\theta_1} e^{i(\omega_1-\omega_2)t} + c.c., \qquad k_2(t) = \kappa_2 e^{i\theta_2} e^{i(\omega_2-\omega_1)t} + c.c.. \tag{S22}$$

The equations of motion are

$$\ddot{u}_1 + \omega_1^2(1-\Delta_1)u_1 - \gamma_1\omega_1(1-u_1^2)\dot{u}_1 + a_1\omega_1^2 u_1^3 = \omega_1^2 \kappa_1(t)u_2,$$
$$\ddot{u}_2 + \omega_2^2(1+\Delta_2)u_2 - \gamma_2\omega_2(1-u_2^2)\dot{u}_2 + a_2\omega_2^2 u_2^3 = \omega_2^2 \kappa_2(t)u_1. \tag{S23}$$

Because two modes have different frequencies, we define different scaling parameters, $\varepsilon_1$ and $\varepsilon_2$, satisfying $\varepsilon_1\omega_1 = \varepsilon_2\omega_2$, i.e., $T = \varepsilon_1\omega_1 t = \varepsilon_2\omega_2 t$. Then we introduce

$$u_1(t) \sim [u_{01}(t)e^{i\omega_1 t} + c.c.], \qquad u_2(t) \sim [u_{02}(t)e^{i\omega_2 t} + c.c.], \tag{S24}$$

and

$$U_1(T) = u_{01}(T/\varepsilon_1\omega_1), \qquad U_2(T) = u_{02}(T/\varepsilon_2\omega_2). \tag{S25}$$

The equations of motion for $U_1(T)$ and $U_2(T)$ are given by

$$i\frac{dU_1}{dT} = +\frac{D_1}{2}U_1 + i\frac{\Gamma_1}{2}(1-|U_1|^2)U_1 - \frac{3\alpha_1}{2}|U_1|^2 U_1 + \frac{K_1 e^{i\theta_1}}{2}U_2,$$
$$i\frac{dU_2}{dT} = -\frac{D_2}{2}U_2 + i\frac{\Gamma_2}{2}(1-|U_2|^2)U_2 - \frac{3\alpha_2}{2}|U_2|^2 U_2 + \frac{K_2 e^{i\theta_2}}{2}U_1. \tag{S26}$$

Here, we define

$$D_n := \frac{\Delta_n}{\varepsilon_n}, \qquad \Gamma_n := \frac{\gamma_n}{\varepsilon_n}, \qquad \alpha_n := \frac{a_n}{\varepsilon_n}, \qquad K_n := \frac{\kappa_n}{\varepsilon_n} \quad (n=1,2). \tag{S27}$$

As in the case of static coupling, the imaginary part leads to

$$|R_1|^2 \sim 1 + \frac{K_1 R_2}{\Gamma_1 R_1}\sin(\varphi_2-\varphi_1-\theta), \qquad |R_2|^2 \sim 1 + \frac{K_2 R_1}{\Gamma_2 R_2}\sin(\varphi_1-\varphi_2+\theta). \tag{S28}$$

Here $\theta = \theta_2 - \theta_1$. From the equations, the phase equations are given by

$$\varphi_1' \sim -\frac{D_1}{2} + \frac{3\alpha_1}{2} + \frac{3\alpha_1 K_1 R_2}{2\Gamma_1 R_1}\sin(\varphi_2-\varphi_1-\theta) - \frac{K_1 R_2}{2R_1}\cos(\varphi_2-\varphi_1-\theta),$$
$$\varphi_2' \sim +\frac{D_2}{2} + \frac{3\alpha_2}{2} + \frac{3\alpha_2 K_2 R_1}{2\Gamma_2 R_2}\sin(\varphi_1-\varphi_2+\theta) - \frac{K_2 R_1}{2R_2}\cos(\varphi_1-\varphi_2+\theta). \tag{S29}$$

By subtracting the first equation from the second equation, we obtain

$$\frac{d\varphi}{dT} = D - C\sin(\varphi-\theta) - E\cos(\varphi-\theta). \tag{S30}$$

Here the parameters are given by

$$D = \frac{D_1 + D_2 + 3(\alpha_2-\alpha_1)}{2}, \quad C = \frac{3\alpha_2 K_2 R_1}{2\Gamma_2 R_2} + \frac{3\alpha_1 K_1 R_2}{2\Gamma_1 R_1}, \quad E = \frac{K_2 R_1}{2R_2} - \frac{K_1 R_2}{2R_1}. \tag{S31}$$

If we approximate $R_1 \sim R_2 \sim 1$,

$$C \sim \frac{3\alpha_2 K_2}{2\Gamma_2} + \frac{3\alpha_1 K_1}{2\Gamma_1}, \quad E = \frac{K_2 - K_1}{2}. \tag{S32}$$

Using the unscaled parameters, we obtain

$$\frac{d\varphi}{dt} = d - c\sin(\varphi - \theta) - e\cos(\varphi - \theta),$$

$$d = \frac{\omega_1 \Delta_1 + \omega_2 \Delta_2 + 3(\omega_2 a_2 - \omega_1 a_1)}{2}, \tag{S33}$$

$$c = \frac{1}{2}\left(\frac{3\omega_2 a_2 \kappa_2}{\gamma_2} + \frac{3\omega_1 a_1 \kappa_1}{\gamma_1}\right), \quad e = \frac{\omega_2 \kappa_2 - \omega_1 \kappa_1}{2}.$$

We obtain the Adler equation with an externally determined phase parameter, $\theta$.

4. Nonlinear parametric coupling

Finally, we investigate the case with nonlinear parametric coupling, where the interaction terms include a higher order one in $u$. For simplicity, we consider the case with third-order nonlinearity.

$$\ddot{u}_1 + \omega_1^2(1-\Delta_1)u_1 - \gamma_1 \omega_1(1-u_1^2)\dot{u}_1 + a_1\omega_1^2 u_1^3 = \omega_1^2 \lambda_1(t) u_1 u_2^2$$
$$\ddot{u}_2 + \omega_2^2(1+\Delta_2)u_2 - \gamma_2 \omega_2(1-u_2^2)\dot{u}_2 + a_2\omega_2^2 u_2^3 = \omega_2^2 \lambda_2(t) u_2 u_1^2 \tag{S34}$$

Here the coupling modulation is given by

$$\lambda_1(t) = \lambda_1 e^{i2(\omega_1-\omega_2)t} + c.c., \quad \lambda_2(t) = \lambda_2 e^{i2(\omega_2-\omega_1)t} + c.c.. \tag{S35}$$

For simplicity, we consider the case that the couplings, $\lambda_1$ and $\lambda_2$ are both real. We obtain

$$\lambda_1(t) u_1(t) u_2(t)^2 \sim \left[\lambda_1 U_1^*(T) [U_2(T)]^2 e^{i\omega_1 t} + c.c.\right],$$
$$\lambda_2(t) u_2(t) u_1(t)^2 \sim \left[\lambda_2 U_2^*(T) [U_1(T)]^2 e^{i\omega_2 t} + c.c.\right]. \tag{S36}$$

Then,

$$i\frac{dU_1}{dT} = +\frac{D_1}{2}U_1 + i\frac{\Gamma_1}{2}(1-|U_1|^2)U_1 - \frac{3\alpha_1}{2}|U_1|^2 U_1 + \frac{\Lambda_1}{2}U_1^*(T)[U_2(T)]^2,$$

$$i\frac{dU_2}{dT} = -\frac{D_2}{2}U_2 + i\frac{\Gamma_2}{2}(1-|U_2|^2)U_2 - \frac{3\alpha_2}{2}|U_2|^2 U_2 + \frac{\Lambda_2}{2}U_2^*(T)[U_1(T)]^2. \tag{S37}$$

As in the case of static coupling, the imaginary part leads to

$$|R_1|^2 \sim 1 + \frac{\Lambda_1 R_2}{\Gamma_1 R_1}\sin 2\varphi, \quad |R_2|^2 \sim 1 - \frac{\Lambda_2 R_1}{\Gamma_2 R_2}\sin 2\varphi. \tag{S38}$$

Here, $\theta = \theta_2 - \theta_1$. Then, the phase equations are given by

$$\varphi_1' \sim -\frac{D_1}{2} + \frac{3\alpha_1}{2} + \frac{3\alpha_1 \Lambda_1 R_2}{2\Gamma_1 R_1}\sin 2\varphi - \frac{\Lambda_1 R_2}{2R_1}\cos 2\varphi,$$

$$\varphi_2' \sim +\frac{D_2}{2} + \frac{3\alpha_2}{2} - \frac{3\alpha_2 \Lambda_2 R_1}{2\Gamma_2 R_2}\sin 2\varphi - \frac{\Lambda_2 R_1}{2R_2}\cos 2\varphi. \tag{S39}$$

As in the case of fundamental harmonics, we finally obtain

$$\frac{d\varphi}{dt} = d - f\sin 2\varphi - g\cos 2\varphi,$$

$$d = \frac{\omega_1 \Delta_1 + \omega_2 \Delta_2 + 3(\omega_2 a_2 - \omega_1 a_1)}{2}, \tag{S40}$$

$$f = \frac{1}{2}\left(\frac{3\omega_2 a_2 \lambda_2}{\gamma_2} + \frac{3\omega_1 a_1 \lambda_1}{\gamma_1}\right), \quad g = \frac{\omega_2 \lambda_2 - \omega_1 \lambda_1}{2}.$$

We obtain the Adler equation of the second order harmonic. This nonlinear and higher harmonic parametric coupling can be implemented in optomechanical systems as shown in [20].

### B. Non-reciprocal time evolution with a finite detuning

We here show the numerical results of the two-tone Adler equation (3) with finite detuning. Typical examples calculated with the parameters, $r = 0.55$ and $\Delta/2\pi = 0.004$ are shown in Fig. S1.

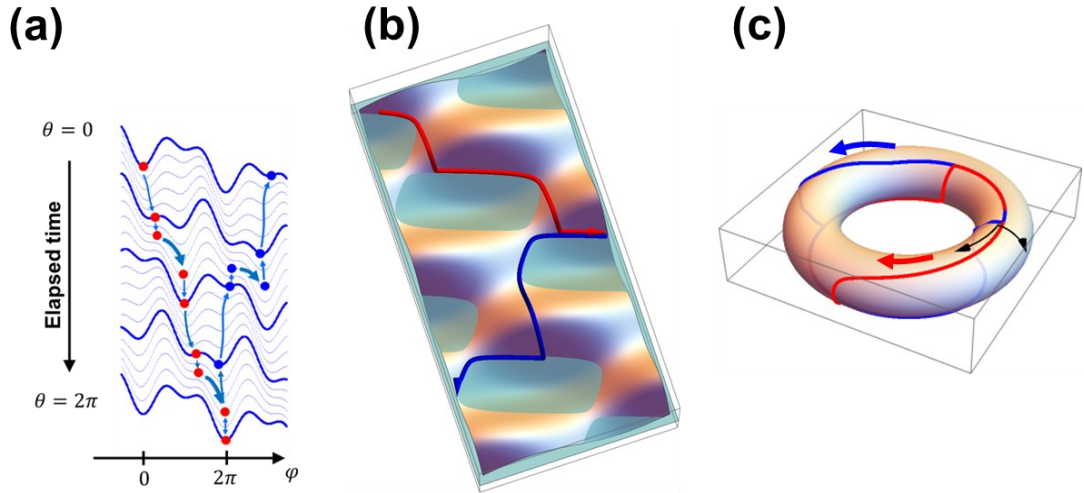

Fig.S1 (a) The time evolution of $\varphi$ calculated by numerically solving the Adler equation shown with the generalized potential. (b) The trajectories superimposed on a bird's-eye view of the force field $f(\varphi, \theta)$ and a horizontal plane ($z(\varphi, \theta) = 0$), by which the negative region of the force field is colored light green and is partially transparent. The trajectory calculated for the reversed motion, i.e., $\theta(t) = -\Omega t$, is also shown as blue in addition to the forward motion, $\theta(t) = \Omega t$ (red). (c) The trajectories are mapped on $\theta - \varphi$ torus surfaces. The used parameters are $\Delta/2\pi = 0.004$, $r = 0.55$. For the calculation, we use $\kappa = 0.1$ and $\Omega/2\pi = 0.0002$ but the result did not show an essential difference if we use a small enough value of $\Omega$.

In contrast to the case with no detuning (Fig.2), the reversed trajectory does not go back to the initial $\varphi$. This is because the value after the phase slip keeps its local minimum even when the drive direction is reversed. The next phase slip is directed to another local minimum that induces non-reciprocal time evolution. From the behavior, we expect that an external large fluctuation applied to the parameter $\theta$ drives the one-way rotation of $\varphi$. The system can be regarded as a "phase ratchet", where the random noise can induce one-way motion of the oscillator phase.

## C. The definition of velocity operator in a closed coordinate space

We performed quantum calculations for the time evolution of the oscillation phase in a generalized Adler equation in the reciprocal space as follows. The particle state is specified by its discrete momentum $k \in \{0, \pm 1, \pm 2, \cdots\}$, and the state of the particle at site $k$ is given by $|k\rangle$, satisfying $\langle k|k'\rangle = \delta_{k,k'}$. The system Hamiltonian is defined by

$$\langle k|\hat{H}|k'\rangle = \frac{k^2}{2m_e}\delta_{k,k'} + v_{k-k'}, \tag{S41}$$

or

$$\hat{H} = \sum_{k=-\infty}^{\infty} \frac{k^2}{2m_e} |k\rangle\langle k| + \sum_{k=-\infty}^{\infty}\sum_{l=-\infty}^{\infty} v_l |k\rangle\langle k-l|. \tag{S42}$$

We construct the phase space as their conjugate one by defining the eigen vector,

$$|\varphi\rangle = \frac{1}{\sqrt{2\pi}} \sum_{k=-\infty}^{\infty} e^{-ik\varphi} |k\rangle. \tag{S43}$$

Note that $\varphi$ is not confined in the space $[0, 2\pi)$ in the definition (S43). It satisfies the periodic boundary condition $|\varphi + 2\pi n\rangle = |\varphi\rangle$ for any integer $n$. The wave function of any state $|\psi\rangle$ in $\varphi$ space can be defined by

$$\psi(\varphi) := \langle \varphi|\psi\rangle = \sum_{k=-\infty}^{\infty} e^{ik\varphi} \langle k|\psi\rangle. \tag{S44}$$

Any wave function satisfies the periodic boundary condition, $\psi(\varphi) = \psi(\varphi + 2\pi n)$. Especially in the case of $|\psi\rangle = |\varphi\rangle$,

$$\langle \varphi|\varphi'\rangle = \frac{1}{2\pi} \sum_{k=-\infty}^{\infty} e^{ik(\varphi-\varphi')} = \sum_{n=-\infty}^{\infty} \delta(\varphi - \varphi' + 2n\pi). \tag{S45}$$

We here explicitly indicate the $2\pi$ periodicity in the delta function.

Then we can introduce the momentum operator, $\hat{p}_\varphi$, which corresponds to the position in 1D lattice space. It is defined by

$$\hat{p}_\varphi = \sum_{k=-\infty}^{\infty} |k\rangle k \langle k|. \tag{S46}$$

It reads

$$\langle k|\hat{p}_\varphi|\psi\rangle = k\langle k|\psi\rangle, \tag{S47}$$

and

$$\langle\varphi|\hat{p}_\varphi|\psi\rangle = \sum_{k=-\infty}^{\infty}\langle\varphi|k\rangle k\langle k|\psi\rangle = \sum_{k=-\infty}^{\infty} e^{ik\varphi}k\langle k|\psi\rangle = -i\frac{\partial}{\partial\varphi}\psi(\varphi). \tag{S48}$$

Then, we must carefully define the phase operator $\hat{\varphi}$ because *it does not satisfy the periodic boundary condition* while keeping $\hat{\varphi}|\varphi\rangle = \varphi|\varphi\rangle$. If we assume $|\varphi\rangle = |\varphi + 2\pi\rangle$, $\hat{\varphi}|\varphi\rangle = \hat{\varphi}|\varphi + 2\pi\rangle = (\varphi + 2\pi)|\varphi + 2\pi\rangle = (\varphi + 2\pi)|\varphi\rangle$, then and it requires $|\varphi\rangle = 0$. We can define $|\varphi\rangle$ but cannot consistently define $\hat{\varphi}$. Instead, we can define the exponential function $e^{i\hat{\varphi}}$ because it is periodic: $e^{i\hat{\varphi}}|\varphi\rangle = e^{i\varphi}|\varphi\rangle$ and $e^{-i\hat{\varphi}}|\varphi\rangle = e^{-i\varphi}|\varphi\rangle$. Then the kinetic term of the Hamiltonian is expressed as

$$\frac{\hat{p}_\varphi^2}{2m_e} + V(\hat{\varphi}), \qquad V(\hat{\varphi}) = \sum_{k=-\infty}^{\infty} e^{ik\hat{\varphi}} v_k, \qquad e^{ik\hat{\varphi}} \equiv \left(e^{i\hat{\varphi}}\right)^k. \tag{S49}$$

Its action on wave function $\psi(\varphi)$ is

$$\langle\varphi|\hat{H}|\psi\rangle = \langle\varphi|\left[\frac{\hat{p}_\varphi^2}{2m_e} + V(\hat{\varphi})\right]|\psi\rangle = -\frac{1}{2m_e}\frac{\partial^2}{\partial\varphi^2}\psi(\varphi) + V(\varphi)\psi(\varphi). \tag{S50}$$

Therefore, we can solve the Schrödinger equation as usual if we limit all the functions to be periodic in $\varphi$. The exponential operators are the raising and lowering operators in reciprocal space as

$$e^{i\hat{\varphi}}|k\rangle = |k+1\rangle, \qquad e^{-i\hat{\varphi}}|k\rangle = |k-1\rangle. \tag{S51}$$

The velocity operator in $\varphi$ space should be also carefully defined. We cannot directly define $\hat{\varphi}$ in contrast to the open-space Thouless pump, but we can use it only in a commutator, for example $\hat{v}_\varphi = d_t\hat{\varphi} = -i[\hat{\varphi},H]$, because $[\hat{\varphi} + 2\pi, H] = [\hat{\varphi}, H]$ satisfies the periodic condition. Here, $d_t$ stands for $d/dt$. Recalling the formula for the angular velocity of a particle in a 2D polar coordinate, $d_t\theta = \sin\theta\, d_t\cos\theta - \cos\theta\, d_t\sin\theta = ie^{i\theta}(d_t e^{-i\theta}) = -ie^{-i\theta}(d_t e^{i\theta})$, we can define the velocity by

$$\hat{v}_\varphi = -\frac{i}{2}e^{-i\hat{\varphi}}(\partial_t e^{i\hat{\varphi}}) + \frac{i}{2}e^{i\hat{\varphi}}(\partial_t e^{-i\hat{\varphi}}),$$
$$= -\frac{1}{2}e^{-i\hat{\varphi}}[e^{i\hat{\varphi}},\hat{H}] + \frac{1}{2}e^{i\hat{\varphi}}[e^{-i\hat{\varphi}},\hat{H}] = \frac{1}{2}\left[e^{-i\hat{\varphi}}\hat{H}e^{i\hat{\varphi}} - e^{i\hat{\varphi}}\hat{H}e^{-i\hat{\varphi}}\right]. \tag{S52}$$

This operator is clearly Hermitian, and the matrix element in $|k\rangle$-basis is

$$\langle k|\hat{v}_\varphi|k'\rangle = \frac{1}{2}\langle k+1|\hat{H}|k'+1\rangle - \frac{1}{2}\langle k-1|\hat{H}|k'-1\rangle = \frac{k}{m_e}\delta_{k,k'} = \langle k|\frac{\hat{p}_\varphi}{m_e}|k'\rangle. \tag{S53}$$

The definition is consistent with the definition of $\hat{p}_\varphi$.

### D. Calculation using Floquet theory of non-adiabatic winding number

The time-dependent Schrödinger equation is given by

$$i\frac{\partial}{\partial t}|\psi(t)\rangle = \hat{H}(\Omega t)|\psi(t)\rangle, \qquad \hat{H}(\Omega t) = \frac{\hat{p}_\varphi^2}{2m_e} - \left[r\cos(\hat{\varphi} - \Omega t) + \frac{1-r}{2}\cos(2\hat{\varphi})\right] \tag{S54}$$

Following the Floquet theory, we assume the form of the solutions

$$|\psi_n(t)\rangle = e^{-i\varepsilon_n t}|\chi_n(\Omega t)\rangle, \qquad |\chi_n(\theta + 2\pi)\rangle = |\chi_n(\theta)\rangle, \tag{S55}$$

where $|\chi_n(\Omega t)\rangle = |\chi_n(\Omega t + 2\pi)\rangle$ is a periodic state vector and $\varepsilon_n$ is the n-th eigen quasi-energy. Then $|\chi_n(\Omega t)\rangle$ satisfies

$$\left[i\frac{\partial}{\partial t} - \widehat{H}(\Omega t)\right]|\chi_n(\Omega t)\rangle = -\varepsilon_n|\chi_n(\Omega t)\rangle \tag{S56}$$

We expand $|\chi_n(\Omega t)\rangle$ and $\widehat{H}(\Omega t) = \widehat{H}(\Omega t + 2\pi)$ by their Fourier series as

$$|\chi_n(\Omega t)\rangle = \sum_{q=-\infty}^{\infty} e^{-iq\Omega t}|\chi_n, q\rangle, \qquad \widehat{H}(\Omega t) = \sum_{q'=-\infty}^{\infty} e^{-iq'\Omega t}\widehat{h}(q'). \tag{S57}$$

From (S56) and (S57), the $k$-space eigen equation for determining $|\chi_n, q\rangle$ and $\varepsilon_n$ becomes,

$$\sum_{k'=-\infty}^{\infty}\sum_{q'=-\infty}^{\infty} \langle k|\widehat{h}(q-q')|k'\rangle\langle k'|\chi_n, q'\rangle - q\Omega\langle k|\chi_n, q\rangle = \varepsilon_n\langle k|\chi_n, q\rangle. \tag{S58}$$

We performed numerical calculations in the $k$ space by using a finite cutoff $|k| \leq k_{max}$. Among a number of identical eigenstates appearing in the calculation [S2], we choose one that well localizes around the small number region of $k$. We confirmed that the value of $\langle k|\chi_n, q\rangle$ is sufficiently suppressed at the edge of calculation region, $k \sim k_{max}$, to confirm that $k_{max}$ is large enough. We used $k_{max}$ from 40 to 200 depending on the value of $\Omega$ and $m_e$. The overall average of operator $\widehat{A}$ is then given by

$$\langle \widehat{A}\rangle = \frac{\Omega}{2\pi}\int_0^{\frac{2\pi}{\Omega}} dt\langle\chi_n(\Omega t)|\widehat{A}|\chi_n(\Omega t)\rangle = \sum_{q=-\infty}^{\infty} \langle\chi_n, q|\widehat{A}|\chi_n, q\rangle. \tag{S59}$$

Especially, $\langle\widehat{H}\rangle$ and $\langle\widehat{v}_\varphi\rangle/2\pi$ provide the average energy and total phase rotation, respectively. The winding number shown in Fig. 5 was calculated as $\langle\widehat{v}_\varphi\rangle/2\pi$ from the formula.

### E. Space-Time reversal

We start from the Hamiltonian,

$$\widehat{H}(\Omega t) = \frac{\widehat{p}_\varphi^2}{2m_e} + \widehat{V}(\Omega t), \qquad \widehat{V}(\Omega t) = -[A_1\cos(\widehat{\varphi} + \Omega t) + A_2\cos(2\widehat{\varphi})] \tag{S60}$$

$\widehat{H}(\Omega t)$ is PT-symmetric because

$$(PT)\widehat{V}(\Omega t)(PT)^\dagger := -[A_1\cos(-\widehat{\varphi} - \Omega t) + A_2\cos(-2\widehat{\varphi})] = \widehat{V}(\Omega t) \tag{S61}$$

Therefore, the PT-reversed time-dependent wave function

$$\mathcal{PT}\psi(\varphi,t) = \sum_{k=-\infty}^{\infty} e^{ik\varphi}\langle k|PT|\psi(t)\rangle := [\psi(-\varphi,-t)]^* = \left[\sum_{k=-\infty}^{\infty} e^{-ik\varphi}\langle k|\psi(-t)\rangle\right]^* \quad \text{(S62)}$$

$$= \sum_{k=-\infty}^{\infty} e^{ik\varphi}\langle k|\psi(-t)\rangle^* = \sum_{k=-\infty}^{\infty} e^{ik\varphi}\langle \psi(-t)|k\rangle$$

also satisfies the time-dependent Schrödinger equation,

$$i\frac{\partial}{\partial t}\mathcal{PT}\psi(\varphi,t) = \left[-\frac{\partial_\varphi^2}{2m_e} + V(\Omega t)\right]\mathcal{PT}\psi(\varphi,t). \quad \text{(S63)}$$

Here, we used

$$\psi(\varphi) = \langle\varphi|\psi\rangle = \sum_{k=-\infty}^{\infty} e^{ik\varphi}\langle k|\psi\rangle. \quad \text{(S64)}$$

The Floquet wave function is defined as

$$\psi_n(\varphi,t) = \langle\varphi|\psi_n(t)\rangle = e^{-i\varepsilon_n t}\langle\varphi|\chi_n(\Omega t)\rangle,$$

$$\langle\varphi|\chi_n(\Omega t)\rangle = \sum_{q=-\infty}^{\infty} e^{-iq\Omega t}\langle\varphi|\chi_n,q\rangle = \sum_{k=-\infty}^{\infty}\sum_{q=-\infty}^{\infty} e^{-iq\Omega t}e^{ik\varphi}\langle k|\chi_n,q\rangle. \quad \text{(S65)}$$

Therefore, the $\mathcal{PT}$-reversed wave function can be expressed as

$$\mathcal{PT}\psi_n(\varphi,t) = [\psi_n(-\varphi,-t)]^* = e^{-i\varepsilon_n t}\sum_{k=-\infty}^{\infty}\sum_{q=-\infty}^{\infty} e^{-iq\Omega t}e^{ik\varphi}\langle k|\chi_n,q\rangle^*. \quad \text{(S66)}$$

Therefore, if the vector components $\langle k|\chi_n,q\rangle$ are all real, $\mathcal{PT}$ is an identical operator, i.e., $\mathcal{PT}\psi_n(\varphi,t) = \psi_n(\varphi,t)$. If we assume $\lambda$ is the eigen value of $\mathcal{PT}$, then $(\mathcal{PT})^2\psi(\varphi,t) = \mathcal{PT}(\lambda\psi(\varphi,t)) = \lambda\lambda^*\psi(\varphi,t) = \psi(\varphi,t)$. Therefore, $\lambda\lambda^* = 1$ so that $\lambda = e^{i\mu}$. The eigen equation is given by

$$\sum_{k'=-\infty}^{\infty}\sum_{q'=-\infty}^{\infty}\langle k|\hat{h}(q-q')|k'\rangle\langle k'|\chi_n,q'\rangle - q\Omega\langle k|\chi_n,q\rangle = \varepsilon_n\langle k|\chi_n,q\rangle \quad \text{(S67)}$$

All the coefficients of $\langle k|\hat{h}(q-q')|k'\rangle$ are real, so that $\langle k|\chi_n,q\rangle$ can be chosen as real. This means that the eigen value of $\mathcal{PT}$ becomes unity. However, we can multiply the wave function by a prefactor of $e^{i\mu}$, where $\mu$ can be any real number, then the eigen value of $\mathcal{PT}$ becomes $e^{2i\mu}$. We can perform the U(1) transformation of wave function as $\psi'_n(\varphi,t) = e^{i\alpha}\psi_n(\varphi,t)$, then $\mathcal{PT}\psi'_n(\varphi,t) = \lambda'\psi'_n(\varphi,t)$ with $\lambda' = e^{-2i\alpha}\lambda$. Therefore, the phase of $\mathcal{PT}$ eigen value does not have any physical meaning. Therefore, the momentum space wave function for the PT-symmetric closed-space Thouless pump can always be chosen as a real function.

References

S1. R. Lifshitz, E. Kenig, and M. C. Cross, "Collective dynamics in arrays of coupled nonlinear resonators" in *Fluctuating Nonlinear Oscillators* ed. by M. I. Dykman (Oxford Univ. Press 2012)

S2. Hideo Sambe, "Steady States and Quasienergies of a Quantum-Mechanical System in an Oscillating Field", Phys. Rev. A 7, 2203 (1973).